%% file: main.tex
\documentclass[runningheads]{llncs}
\usepackage[T1]{fontenc}
\usepackage{amsmath,amssymb,mathtools,booktabs,longtable,array,tabularx,graphicx}
\makeatletter

\let\c@theorem\relax

\makeatother
\usepackage{amsthm}
\newtheorem{theorem}{Theorem}
\newtheorem{lemma}[theorem]{Lemma}
\newtheorem{corollary}[theorem]{Corollary}
\newtheorem{proposition}[theorem]{Proposition}
\theoremstyle{definition}
\newtheorem{definition}[theorem]{Definition}
\newtheorem{hypothesis}{Hypothesis}
\usepackage{tikz,float}
\usetikzlibrary{positioning,arrows.meta}
\usepackage[hidelinks]{hyperref}
\newcommand{\poly}{\operatorname{poly}}

\newcommand{\Top}{\operatorname{Top}}

\title{Counting Feedback Vertex Sets of All Sizes in Subexponential Time, Conditionally}
\titlerunning{Counting Feedback Vertex Sets of All Sizes, Conditionally}
\author{Mengyang Wu \and Minghao Wang}
\authorrunning{Mengyang Wu and Minghao Wang}
\institute{}
\begin{document}
\raggedbottom
\maketitle
\begin{abstract}
Assuming a Boolean tensor contraction hypothesis, we study exact feedback vertex set counting in general graphs.
For an $n$-vertex, $m$-edge simple graph, we compute the complete size distribution deterministically in
$\poly(n+m)2^{O(n(\log\log(n+4))^2/\sqrt{\log(n+2)})}$ time and polynomial working space.
The algorithm combines a whole-edge forest identity, subset-index scalar networks, sparse parity filtering, and disjoint high-degree branching.
A binomial inversion recovers the multiplicities introduced by exceptional-set guesses.
The counting interface supports fixed vertices, minimum solutions, and unranking.
Component markers yield induced-tree and induced-path counts within the same bound.
Sparse Hamilton, forest, and Steiner edge-tree counting follow from the scalar encoding.
All filtering queries, coefficient precision, and recovery costs are included in the stated time and space bounds.
\end{abstract}
\section{Introduction}
\subsection{Problem and motivation}
A feedback vertex set removes all cycles of a graph. Counting these sets by size describes the full feasible-size distribution and supports minimum-solution recovery and unranking. The global constraint is acyclicity of the retained induced graph; arbitrary input density makes a direct incidence encoding expensive.
\subsection{Main results}
Let $G$ be a simple undirected graph with $n$ vertices and $m$ edges. A selector pin requires an original vertex to be retained or deleted; unranking returns the deletion set with a specified rank in a fixed vertex order. The tensor-contraction premise is stated precisely in Hypothesis~\ref{hyp:contraction}.
\begin{theorem}[{General-graph exact FVS counting}]
\label{thm:general-fvs}
Assuming the Hypothesis~\ref{hyp:contraction}, every simple undirected
$n$-vertex graph admits deterministic computation of all feedback
vertex set cardinality counts in
\begin{equation}
 \poly(n+m)\,2^{O\left(
       \frac{n(\log\log(n+4))^2}{\sqrt{\log(n+2)}}\right)}
                                                               \label{eq:formula-28}
\end{equation}
time and polynomial working space, with no density restriction.
The same bound permits arbitrary original selector pins, a minimum
feedback vertex set, and unranking a deletion set of a given size.
Enumeration has this delay and polynomial working space, apart from
writing the output. 
\end{theorem}
\begin{table}[tbp]\centering\small
\begin{tabularx}{\linewidth}{@{}p{.24\linewidth}X@{}}\toprule
Source / task & Bound and scope \\
\midrule
Li--Nederlof~\cite{linfvs} & Randomized minimum FVS in $O^*(1.6297^n)$ time. \\
Cut\&Count~\cite{cutcount} & One-sided Monte Carlo FVS decision in $3^{\mathrm{tw}}n^{O(1)}$ time on a tree decomposition. \\
This paper & All cardinality counts in $\poly(n+m)2^{O(n(\log\log(n+4))^2/\sqrt{\log(n+2)})}$ time; deterministic, polynomial space, under Hypothesis~\ref{hyp:contraction}. \\
\bottomrule\end{tabularx}
\caption{Representative FVS bounds, with their output and assumptions. The parameters and tasks are stated explicitly.}
\end{table}

The density-sensitive oracle is Theorem~\ref{thm:fvs}. Component markers give Theorem~\ref{thm:components}, including induced trees and paths. Sparse Hamilton, forest, and Steiner edge-tree consequences are collected in Section~\ref{sec:companions}.
\subsection{Technical overview}
\input{technical_overview_body}

\subsection{Related work}
Bodlaender, Cygan, Kratsch, and Nederlof~\cite{bckn} give deterministic integer counting for Hamilton cycles in $\widetilde O(15^{\mathrm{tw}}\mathrm{tw}^{O(1)}n^2)$ and Steiner trees in $\widetilde O(10^{\mathrm{tw}}\mathrm{tw}^{O(1)}n^3)$. Their structural parameter is treewidth. The FVS comparisons in Table~1 distinguish decision, optimization, and complete counting.

Claude (Anthropic)~\cite[Theorem 1]{anthropic2026} gives a deterministic
CNF counting bound with an absolute constant in
$2^{O(\min\{k^2n\log\log n/\sqrt{\log n},km/\sqrt{\log m}\})}$.
Our scalar encoding uses the ring-valued tensor interface of Theorem D.1.

The algebraic cancellation identities~\cite{forest,staples}, sparse parity
framework~\cite{base}, and induced-sparse-subgraph branching
idea~\cite{inducedsparse} provide the starting tools. Our contributions are
the whole-edge scalar interface with complete wire and precision budgets,
polynomial-query parity selection, and the disjoint branching with exact
multiplicity recovery that together yield the all-density theorem.

\subsection{Organization}
Section~\ref{sec:prelim} gives the contraction premise and forest identity. Section~\ref{sec:framework} states the complete algorithm and its dependencies. Sections~\ref{sec:sparse-global} and~\ref{sec:forest-parity} build its density-sensitive oracle; Section~\ref{sec:dense-fvs} proves the all-density theorem. The remaining sections give component and sparse companion counts. The appendices contain the detailed recovery and auxiliary constructions.
\section{Preliminaries and contraction hypothesis}\label{sec:prelim}
Graphs are simple and undirected unless arcs are specified. An FVS is a
set $D$ for which $G[V\setminus D]$ is a forest. A singleton is a tree and
a path; the empty forest has zero components. All logarithms are base two
except where a natural logarithm is written. The notation $\poly$ denotes
a fixed polynomial factor.

\begin{hypothesis}[{Theorem D.1 of OpenAI~\cite{base}}]
\label{hyp:contraction}
Let $s\ge2$ and let a bipartite Boolean tensor network have $O(s)$ wires,
polynomially many vertices, maximum degree at most $\lceil s^{9/10}\rceil$,
and description length $\mathcal L$. Its entries lie in
$(\mathbb Z/2^{P_0}\mathbb Z)[V]/(V^\ell)$, with $P_0,\ell=O(s)$.
Suppose a local entry is computable in
$(s+\mathcal L)^{O(1)}2^{O(s^{9/10})}$ time and polynomial space.
Then its contraction is computable deterministically in
$(s+\mathcal L)^{O(1)}2^{O(s/\sqrt{\log s})}$ time and polynomial space.
\end{hypothesis}

\begin{theorem}[{Sparse parity checks, from~\cite[Lemma 6.1]{base}}]
\label{thm:sparse-code}
For positive integers $m,H$, put $B=\max(1,\lceil m/H\rceil)$,
$\mu=\lfloor\log_2m\rfloor+1$, $P=20000B^2\mu^2$, and
$D_0=320000B^2\mu^3$. There is a binary matrix with $m$ columns,
at most $3H$ rows of support at most $D_0$, and no nonzero kernel word
of weight less than $m/P$.
\end{theorem}
The local codes and shifts needed by this construction are made explicit
below and in Appendix~\ref{app:shifts}. The imported sparse-check statement
and Hypothesis~\ref{hyp:contraction} are used at their respective size parameters.

\subsection{The forest identity}\label{sec:forest-identity}
For forests, use the $2n$ Grassmann atoms
$\bar\psi_1,\psi_1,\ldots,\bar\psi_n,\psi_n$, set
$\tau_v=\bar\psi_v\psi_v$, and let $\Top$ extract the coefficient of
$\tau_1\cdots\tau_n$ in canonical order. The imported forest identity is
\begin{equation}
 \Top\prod_v(1+\tau_v)\prod_{uv\in E}(1+UQ_{uv})
       =\sum_{F\text{ forest}}U^{|F|},                \label{eq:forest}
\end{equation}
where $Q_{uv}=(\bar\psi_u-\bar\psi_v)(\psi_u-\psi_v)-\tau_u\tau_v$.
We replace each edge factor by the product of its five elementary
monomial factors, and each vertex factor by its single mass factor.
This is an integer identity: the even monomials commute, each has
square zero, and all products of two distinct summands within
$Q_{uv}$ cancel, since its quadratic part is decomposable and its
quartic part overlaps it. Equivalently, the product is the finite
exponential of $UQ_{uv}$, which equals $1+UQ_{uv}$.
Each elementary term has two or four physical atoms and grade one.
There are $O(n+m)$ ports. The degree in $U$ of
each selected-term contribution is at most $n$; cancellation gives
the usual forest degree at most $n-1$.

\section{Algorithm and proof framework}\label{sec:framework}
The algorithm for Theorem~\ref{thm:general-fvs} has an outer combinatorial reduction and an inner algebraic oracle. Its complete sequence is as follows.
\begin{enumerate}
\item Choose the thresholds $t,D$ of Section~\ref{sec:dense-fvs} and enumerate exceptional retained sets $B$ of size at most $b$ from~\eqref{eq:formula-22}.
\item For each $B$, branch on high-current-degree vertices outside $B$, specifying the exact retained neighbours in a retaining branch. The disjoint-leaf lemma in Section~\ref{sec:dense-fvs} bounds the number and density of the leaves.
\item Count each leaf by Theorem~\ref{thm:fvs}, with its original pins and degree caps. The scalar-network lemmas of Section~\ref{sec:sparse-global}, exact separation in Lemma~\ref{lem:separation}, and weighted oracle in Lemma~\ref{lem:moment} supply this call.
\item Accumulate counts by retained size and exceptional-set size, then apply~\eqref{eq:formula-27}. Reverse the retained-size index to obtain every FVS cardinality count.
\end{enumerate}
The correctness proof has two independent multiplicity arguments. The algebraic oracle cancels auxiliary expansions inside each graph object, and the outer binomial inversion corrects overlap among exceptional-set guesses. The proof in Section~\ref{sec:dense-fvs} combines their costs and derives the final exponent. All calls are streamed to preserve polynomial space.

\section{Selected-port compression with ordinary scalar tensors}
\label{sec:sparse-global}

Direct incidence networks for Hamilton cycles and forests use $O(n+m)$ physical wires.
We encode selected ports and factor the signs into ordinary scalar tensors. We put
\begin{equation}
 s=n\bigl(1+\log_2(1+m/n)\bigr),\qquad n\geq2.       \label{eq:formula-1}
\end{equation}
All graphs in this section are simple, so $m=O(n^2)$.
\subsection{An ordinary network for signed covers of atoms}
We use $2n$ ordered physical atoms, in adjacent pairs belonging to vertices.
A term chooses an even-sized support of at most four atoms, a scalar
coefficient, and a grade in $\{0,1\}$. Grade-one atoms contribute
the exterior inversion sign; grade-zero atoms contribute no such sign.
Every selected term has even grade, so the sign is invariant under reordering selected terms. Every physical atom must be covered exactly
once. Local conditions may require the two atoms of a vertex to have
the same grade, or forbid grade one at a specified vertex.

For an optional term, assign its factor to the group of its smallest
physical atom. We partition vertices into $J=\lceil n^{1/4}\rceil$
consecutive groups, each of size at most $\lceil n/J\rceil$.
The atom groups on the other side use the same partition.
A bounded number of positive classical predicates can additionally be
attached to a selected term; these receive their own column groups.
For now ignore those classical ports.

We give each occurrence of an atom in a term its own port. A port belongs
to a factor group $g$, an atom group $h$, and a grade $c$.
Let $N_{ghc}$ be the number of possible ports in that bucket.
We enumerate their selected counts $d_{ghc}$. Only $2n$ physical ports
are selected, and each atom column has its prescribed total size.
For a fixed count array, use
\[
 \left\lceil\log_2\binom{N_{ghc}}{d_{ghc}}\right\rceil
\]
ordinary Boolean wires to index the selected subset, with unused binary
codes assigned entry zero. Lexicographic subset decoding uses exact
binomial coefficients of polynomial bit length, as in source Appendix D.3.

The factor tensor checks that either every port of a term or none is
selected. It multiplies the selected coefficients and the inversion
sign of its own grade-one atoms, with terms in a fixed local order.
The atom tensor checks exact coverage, the local grade conditions,
and the inversion sign between grade-one atoms coming from different
factor groups but lying in this atom group. Its ports specify both
the atom identities and the supplying factor groups.

\begin{lemma}[{Complete scalar sign factorization}]
For a selected-port count array $d_{ghc}$, let the local factor signs count inversions within each factor group and the local atom signs count inversions from distinct factor groups in the same atom group. Their product, multiplied by
\begin{equation}
 (-1)^{\chi(d)},\qquad
 \chi(d)=\sum_{g<g',\ h>h'}d_{gh1}d_{g'h'1}\pmod2,   \label{eq:formula-2}
\end{equation}
is exactly the exterior sign of the full selected grade-one sequence.
Consequently summing the ordinary contractions over the count arrays
equals the signed cover sum, with multiplicity one for every selection.
\end{lemma}
\begin{proof}
The proof partitions inversions according to their factor and atom groups. The terms are ordered by factor group and then by their local order.
An inversion either has both atoms in one factor group, or has them
in distinct factor groups. The former is counted by that factor tensor.
In the latter case, equal atom groups are handled by the corresponding
atom tensor, while distinct atom groups contribute precisely \eqref{eq:formula-2}.
These three cases partition all inversions. Grade-zero and classical
ports contribute none. Every selected port set has one count array and
one valid subset index per bucket; all-or-none and coverage checks
give the converse correspondence.
\end{proof}

\begin{lemma}[{Wire and degree budgets}]
Consider the subset-index construction of Section~\ref{sec:sparse-global} restricted to factor rows and physical-atom columns, with classical predicate ports omitted. Put $s=n(1+\log_2(1+m/n))$. Suppose the input has $I=O(n+m)$ ports, $O(n)$ selected ports, and selected terms of bounded support. The signed cover sum is a sum of
$2^{O(\sqrt n\log(n+2))}$ ordinary networks, each with $O(s)$ wires,
maximum degree $O(n^{3/4}\log(n+2))$, and polynomial-time local entries.
The descriptions and their enumeration use polynomial space.
\end{lemma}
\begin{proof}
The proof bounds subset-index length by entropy and then controls the load at each tensor. There are $O(J^2)$ buckets. With $K=O(n)$ selected ports, source
Appendix D.3's product bound gives
\begin{equation}
 M\leq O(J^2)+\log_2\binom{I+K}{K}
   \leq O(J^2)+K\log_2\frac{e(I+K)}K=O(s).          \label{eq:formula-3}
\end{equation}
The arrays have at most $(K+1)^{O(J^2)}$ possibilities.
Every selected optional term consumes its owner atom, so a factor
group can contain only $O(n/J)$ selected terms, even if many input
terms have that owner. We discard arrays whose row sum exceeds the
corresponding bounded-support cap. Each atom column has $O(n/J)$
selected ports as well. Therefore the wire degree of either vertex is
at most $O(J+(n/J)\log(I+2))$, which is the stated bound for simple
graphs. This selected-load argument is essential: raw incidence
degree can be much larger. Each local tensor decodes polynomially
many bits, performs the explicit coverage and sign checks, and
multiplies polynomial-bit coefficients. No local permanent or
exponential subset sum is hidden in its entry procedure.
\end{proof}

We will also use at most $n$ required factors, each a sum of a fixed
number of even monomials. We put these factors in $J$ additional rows,
balancing their \emph{number}. A row checks that exactly one alternative
is selected per required factor. Its selected load is again $O(n/J)$;
the number of buckets and all bounds above change only by constants.

\section{Whole-edge parity filtering for induced forests}
\label{sec:forest-parity}

Whole-edge indicators retain the edge factor selected in the forest
identity. Sparse parity checks then separate the induced forests from all
forest edge sets on selected vertices.

Edgeless graphs are handled directly. For compatible retained and deleted
vertex pins, let $d_0$ be the number of vertices pinned deleted and $f$ the
number of unpinned vertices. Every choice of the free vertices induces a
forest, so the deletion-size generating polynomial is
\[
 V^{d_0}(1+V)^f.
\]
Its coefficients are computed in polynomial time; nonnegative upper bounds
on retained degrees are automatically satisfied. This base case also covers
empty graphs, single-vertex graphs, and edgeless residual instances.

For the remaining construction, let $G$ be a simple graph with $n\geq2$
vertices and $m\geq1$ edges. We set
\begin{equation}
 L=1+\log_2(1+m/n),\qquad
 H=n+\lfloor\log_2\tbinom{m+n}{n}\rfloor+1,
 \qquad p=H+1.                                      \label{eq:formula-4}
\end{equation}
In particular $H=O(nL)$ and $H\geq n+1$. We consider the finite object set
\begin{equation}
 \Omega=\{(S,F):S\subseteq V(G),\ F\subseteq E(G[S]),\ F\text{ acyclic}\}.
                                                               \label{eq:formula-5}
\end{equation}
We write $z_v=1_{v\in S}$ and $f_e=1_{e\in F}$. Its violation coordinate is
$v_e=z_uz_v+f_e\pmod2$ for $e=uv$. Thus $v=0$ exactly when $F=E(G[S])$.
Every induced forest has exactly one zero-violation object. Also
\begin{equation}
 |\Omega|\leq2^n\sum_{j=0}^{n}\binom mj
             \leq2^n\binom{m+n}n<2^H.              \label{eq:formula-6}
\end{equation}
The second inequality follows by the Vandermonde convolution with
$\binom nj\geq1$. All counts refer to the objects in \eqref{eq:formula-5}.

\subsection{The parity filter and object-mass bound}
For a binary matrix $A$ with $m$ columns, let
\begin{equation}
 N_A(V)=\sum_{(S,F)\in\Omega:\ Av(S,F)=0}V^{n-|S|},
 \qquad N(A)=N_A(1).                                \label{eq:formula-7}
\end{equation}
Every coefficient is nonnegative and $N(A)<2^H$.
We apply Theorem~\ref{thm:sparse-code} with the object-mass parameter $H$ and
\begin{equation}
\begin{aligned} B&=\max(1,\lceil m/H\rceil),&\mu&=\lfloor\log_2m\rfloor+1,\\ P&=20000B^2\mu^2,&D_0&=320000B^2\mu^3.\end{aligned}\label{eq:formula-8}
\end{equation}
The resulting matrix has the row, support, and distance bounds of Theorem~\ref{thm:sparse-code}.

The local code has the following explicit construction. In its notation take
$d=\min(m,D_0)$, $\delta=1/(100B\mu)$, and
$h=\lceil\delta d\rceil-1$. If $h=0$ use no local checks. Otherwise let
$r=\lceil\log_2d\rceil$, choose distinct
$\alpha_1,\ldots,\alpha_d\in\mathbb F_{2^r}$, and impose
\begin{equation}
 \sum_{j=1}^{d}\alpha_j^k x_j=0
       \quad\text{in }\mathbb F_{2^r},\qquad 0\leq k<h.           \label{eq:formula-9}
\end{equation}
We expand each field equation into $r$ binary equations. Any at most $h$
columns are independent: their first as many field rows form a square
Vandermonde matrix with nonzero determinant. Thus a nonzero binary kernel
word has weight at least $h+1\geq\delta d$. The binary row count satisfies
$hr<d/(100B)\leq\lceil d/(8B)\rceil$ because $r\leq\mu$.

A field presentation can be found by enumerating irreducible binary
polynomials of degree $r$; even trial factoring costs polynomial time in
$2^r\leq2d$. Local-code construction is therefore polynomial in $m$.

Lemma~\ref{lem:bias} constructs the required shifts in polynomial time;
its exact conditional-expectation proof is in Appendix~\ref{app:shifts}.
Together with the displayed Vandermonde construction this gives
\begin{equation}\poly(n+m)\quad\text{time and space}.\label{eq:formula-10}\end{equation}

\begin{lemma}[{Exact separation}]\label{lem:separation}
Let $N_A(V)$ be the object polynomial in~\eqref{eq:formula-7}, with $H,P$ as in~\eqref{eq:formula-4} and~\eqref{eq:formula-8}, and with the initial sparse-check matrix of Theorem~\ref{thm:sparse-code}. At most $3H$ additional parity rows, each of support at most $P$, suffice
to obtain
\begin{equation}
 N_A(V)=\sum_{S:\,G[S]\text{ acyclic}}V^{n-|S|}.\label{eq:formula-11}
\end{equation}
The same matrix separates every subfamily obtained by pinning vertices or edges.
\end{lemma}
\begin{proof}
The proof appends parity rows that remove a constant fraction of the surviving violations. A random list $((e_1,c_1),\ldots,(e_P,c_P))$, with uniform edge coordinates
and independent coefficient bits, hits the support of each surviving
nonzero violation with probability greater than $1/2$. Conditional on a
hit, its row parity is uniform. It therefore rejects each such object with
probability at least $1/4$. If $D(A)$ is their surviving number, a
minimum-count row leaves at most $3D(A)/4$. The zero-violation count is
independent of the row. After $3H$ rows the remaining number is less than
$(3/4)^{3H}2^H=(27/32)^H<1$, hence zero. Elimination is objectwise and
therefore holds for every restricted subfamily and every coefficient.
Appendix~\ref{sec:conditional} constructs these rows using at most
$6HP_*m$ weighted queries, where $P_* =\min(P,m)$.
\end{proof}
\subsection{Coefficientwise cancellation with whole-edge events}
We use the multivariate version of the forest identity \eqref{eq:forest}, originating in
Caracciolo et al.~\cite{forest}, equation (13) at $t=1,u=-1$.
We give each edge factor $1+w_eQ_e$ the six states: empty, or one of the
five canonical monomials of $w_eQ_e$. Selected nonempty states are
mutually exclusive \emph{even when their atom supports are disjoint}.
An additional classical edge-event port records whether the edge is
selected. All selected edge states require $z_u=z_v=1$.

For a fixed edge set $F$, the signed sum of these choices and of vertex
mass choices is the coefficient of $\prod_{e\in F}w_e$ in the
multivariate identity. It is one when $F$ is acyclic and zero otherwise.
This follows coefficientwise, since the original identity allows
independent formal weights $w_e$. Multiplication by any function of
$z$ and the edge-event vector $f$ preserves the coefficientwise
cancellation. In particular it permits the parity indicators defining
\eqref{eq:formula-7}, arbitrary pins of $z$ or $f$, and the deletion weight.

\subsection{Integral parity lift and its total order budget}
A row of $A$ gives a parity of terms $f_e$ and $z_uz_v$.
Its support has size at most $D_0$ or $P$, so it has at most
\begin{equation}
 \tau=2\max(D_0,P)=O(B^2\mu^3)                     \label{eq:formula-12}
\end{equation}
monomials, each of degree at most two. After any pins, remove zero
terms, move constants to the right side, and cancel duplicate terms
over $\mathbb F_2$. The surviving row has the form
$\bigoplus_j t_{ij}=b_i$, where every $t_{ij}$ is a positive predicate
$f_e$, $z_v$, or $z_uz_v$. Its integer indicator is
\begin{equation}
 R_i=\frac{1+(-1)^{b_i}\prod_j(1-2t_{ij})}{2}.
                                                               \label{eq:formula-40}
\end{equation}
The empty selection has coefficient $1-b_i$; a nonempty selection of
$k_i$ distinct terms has coefficient $(-1)^{b_i+k_i}2^{k_i-1}$.
These are integer coefficients; there is no division by two inside
$\mathbb Z/2^p\mathbb Z$.

Let $a\leq6H$ be the number of parity rows. A product selection of total
order $T=\sum_i k_i$ has valuation at least $T-a$. Therefore retain only
\begin{equation}
 T\leq K_0=p+a-1=H+a.                              \label{eq:formula-41}
\end{equation}
All other terms vanish coefficientwise modulo $2^p$, before any
Grassmann cancellation or summation over selectors. For a parity
predicate $f_e$, an edge column requires the whole-edge event to be
selected.
For $z_v$ or $z_uz_v$, vertex columns require the indicated selectors.
Repeated positive requirements are harmless since the variables are
Boolean. Each selected parity term has one or two classical ports.

\subsection{Explicit ordinary network and degree control}
We take $J=\lceil H^{1/4}\rceil$ groups. There are $J$ graph-factor rows,
$J$ parity-factor rows, $J$ physical-atom columns, $J$ vertex-selector
columns, and $J$ edge-event columns. Assign every selected graph
monomial to the group containing its smallest physical atom. We partition
the $a$ parity rows into $J$ groups with at most $\lceil a/J\rceil$
rows each. Within the two classical column families, balance the input
incidence counts of individual variables greedily.

As before, make a separate port for each physical or classical
occurrence, and index selected port subsets per group pair and grade.
We enumerate the bucket counts $d_{ghc}$ and additionally the total parity
order in each parity group, requiring their sum at most $K_0$.
The order of a selection is its number of monomials.
The graph-factor entries check all-or-none selection of each monomial
and multiply its coefficient. Parity-factor entries additionally check
their public orders and multiply the row coefficients in \eqref{eq:formula-40}.
Physical columns require exact atom coverage. Edge columns check
that each edge has at most one event, exactly one if pinned to one,
no event if pinned to zero, and one whenever a parity term requires it.
Vertex columns contribute $1$ if the vertex is required to be retained,
and $1+V$ if it is unpinned and unrequired. A pinned deleted vertex
contributes $V$. Thus a local entry is computable by explicit subset
decoding, incidence checks, and polynomial arithmetic.

All physical monomials have even degree. Their complete sign is given
by the two local inversion signs and the public phase \eqref{eq:formula-2}.
Classical ports contribute no sign. For every surviving selection there
is exactly one count/order array and one set of valid subset indices.
Conversely, nonzero entries force precisely an allowed macro/lift
selection. Hence the sum of these ordinary contractions is \eqref{eq:formula-7}
modulo $2^p$ in the ring
\begin{equation}
 (\mathbb Z/2^p\mathbb Z)[V]/(V^{n+1}).             \label{eq:formula-42}
\end{equation}

There are exactly $2n$ selected physical ports, at most $2n$ graph
selector ports, at most $n$ edge-event ports, and at most $2K_0$ parity
predicate ports. We put $K=5n+2K_0=O(H)$. The input has
$I=O(n+m+a\tau)$ ports. The subset entropy bound gives
\begin{equation}
 M\leq O(J^2)+K\log_2\frac{e(I+K)}K
   =O\left(H[1+\log(1+I/H)]\right).                \label{eq:formula-43}
\end{equation}
There are at most $(K+1)^{O(J^2)}$ public arrays, including the
$J$ order counters. They can be streamed in polynomial space.

For the degree argument first suppose $m\leq n^{33/32}$. In the
unrestricted oracle network, pin every $z_v$ or $f_e$ with more than
$H^{3/4}$ input incidences, in all ways. There are
$O(I/H^{3/4})$ such variables. Each branch restricts the objects in
\eqref{eq:formula-5}, so the branches sum without multiplicity. A pinned $f_e=1$
requires the macro event, and its parity occurrences disappear upon
normalization; a pinned $f_e=0$ deletes its macro alternatives. An
inconsistent edge and vertex pin gives zero. After normalization, the remaining unpinned variables
have incidence at most $H^{3/4}$. A fixed edge retains only its at
most five event ports. Constants and scalar vertex weights cause no
additional high-degree node.

The graph row's selected load is $O(n/J)$: each selected graph
monomial consumes its owner atom. Its wire degree is at most
$O(J+(n/J)\log(I+2))$. The same bound holds for physical columns.
Each parity row group has raw load $O((a/J+1)\tau)$; each balanced
classical column has raw load $O(I/J+H^{3/4}+1)$.
For any bucket, $\lceil\log_2\binom Nd\rceil\leq N$,
so raw load bounds wire degree where needed. Using
$B=O(n^{1/32})$, $\mu=O(\log n)$, $H=O(n\log n)$ gives
\begin{equation}
 \begin{split}
 I&=O(n^{17/16}(\log(n+2))^4),\\
 \#\text{pinned variables}&=O(n^{5/16}(\log(n+2))^4),\\
 \Delta&=O(n^{13/16}(\log(n+2))^4).
 \end{split}                                                    \label{eq:formula-44}
\end{equation}
These loose polynomial logarithmic factors suffice: $13/16<9/10$.
For sufficiently large $n$ the maximum degree is below the source
threshold at the root scale below. Bounded exceptional input sizes
can be handled by direct enumeration.

The weighted moments used to choose the separating rows have the same
root scale. Lemma~\ref{lem:moment}, proved in
Appendix~\ref{sec:conditional}, includes their precision, incidence,
and local-entry costs.

\begin{theorem}[{Density-sensitive complete feedback vertex counting}]
\label{thm:fvs}
Assuming Hypothesis~\ref{hyp:contraction}, all feedback vertex set cardinality
counts of a simple $n$-vertex, $m$-edge graph can be computed
deterministically in polynomial space and time
\begin{equation}
 \poly(n+m)2^{O\left(
 \frac{nL[L+\log\log(n+4)]}{\sqrt{\log(n+2)}}\right)},
 \qquad L=1+\log_2(1+m/n).                          \label{eq:formula-45}
\end{equation}
The same bound includes a minimum deletion set and unranking any
deletion set of specified cardinality. Direct $2^n\poly(n+m)$ enumeration
is an additional alternative, so one may take the faster of these two bounds.
\end{theorem}
\begin{proof}
The proof checks the contraction hypotheses at the compressed scale and then recovers the integer coefficients. Edgeless instances use the direct polynomial at the start of this section, including pinned calls during recovery. Assume henceforth that $m\geq1$. For $m\leq n^{33/32}$, \eqref{eq:formula-43} is bounded by a constant times
\begin{equation}
 s=nL[L+\log\log(n+4)],                            \label{eq:formula-46}
\end{equation}
since $\log(1+I/H)=O(L+\log\log(n+4))$. We choose an integer scale that
is a sufficiently large constant multiple of \eqref{eq:formula-46}. The network
has $O(s)$ wires, maximum degree at most $s^{9/10}$, polynomial-size
descriptions and polynomial-time local entries. The coefficient
precision $p=H+1$ and truncation length $n+1$ are both $O(s)$.
Hypothesis~\ref{hyp:contraction} applies to \eqref{eq:formula-42}. For these unweighted
calls, arrays and pin branches are enumerated sequentially. Their exponents,
$O(\sqrt H\log(H+2))$ and $O(n^{5/16}\log^4(n+2))$, are absorbed
in $O(s/\sqrt{\log(n+2)})$.

The initial code cost \eqref{eq:formula-10} is polynomial. We use the polynomial-query
selection of Section~\ref{sec:conditional}; Lemma~\ref{lem:moment}
charges every weighted moment at the same scale, including discarded
candidates. 
For the unweighted calls, counts used in comparisons are exact integers:
the residue of $N(A)$ is less than $2^H<2^p$ by \eqref{eq:formula-6}. After
separation a final polynomial-valued count gives \eqref{eq:formula-11}, coefficientwise
without an interpolation loss. Its least nonnegative coefficients
are the desired integers.

If $m>n^{33/32}$, then $L>\log_2(n)/32$ and the exponent in
\eqref{eq:formula-45} dominates $n$ for sufficiently large $n$. Direct enumeration
of vertex sets and a forest test therefore proves the same displayed
bound throughout the dense range. 

For recovery, find the least nonzero cardinality coefficient. Process
vertices in order and count that coefficient with the next vertex
pinned deleted. We retain that branch when nonzero, or use the other
branch; for unranking subtract its count when proceeding to the
other branch. The same completed separating matrix works for all
prefix restrictions by the lemma. There are $O(n)$ further oracle
calls, with the original budgets retained as upper bounds. Computing
arbitrary-rank solutions at a prescribed cardinality is identical.
Streaming successive ranks gives enumeration with this delay and
polynomial working space, in addition to unavoidable output cost.
\end{proof}

\section{All-density feedback vertex counting by disjoint sparsification}
\label{sec:dense-fvs}

Induced forests admit a disjoint sparsification because their selected
graphs have small average degree. High-degree guessing and branching are
used for induced sparse-subgraph optimization by Novotn\'a et
al.~\cite{inducedsparse}. The counting reduction below fixes every selected
neighbor in a retaining branch, imposes residual degree caps, and recovers
the multiplicity of exceptional-set guesses by binomial inversion.

\subsection{Degree-capped forest objects}
Let $R$ be a graph on $N$ vertices with a set $B$ pinned retained.
We require $\deg_{R[S]}(v)\leq t$ for every retained $v\notin B$.
In the intermediate object set \eqref{eq:formula-5}, impose instead
\begin{equation}
 \deg_F(v)\leq t\qquad(v\notin B).                 \label{eq:formula-20}
\end{equation}
After separation $F=E(R[S])$, so \eqref{eq:formula-20} is exactly the required
induced-degree condition. The object set is still nonnegative, still
has the bound \eqref{eq:formula-6}, and the source greedy separation applies to
this restricted family without change.

The network implementation counts whole-edge incidences at the
vertex columns. Mark a macro-edge selector port as a degree event,
distinguishing it from positive selector requirements coming from a
parity monomial. A selected macro edge contributes one event at each
endpoint even if its Grassmann monomial uses atoms at only one endpoint.
The column tensor assigns zero to an assignment exceeding its cap. For a capped vertex
already pinned to one, keep these event ports; the degree event remains after the selector is pinned. Coefficientwise
cancellation for a fixed $F$ remains valid, because its degree is
independent of which of the five macro monomials was selected.

We use this extension only at leaves where every capped vertex has
input degree at most $D$, with $D$ a fixed power of the logarithm of
the original size. The extension adds no ports beyond the existing
at most two graph-selector ports per macro edge. After high-incidence
selector pins, at most $5D$ graph-event occurrences remain at a
capped vertex; its parity occurrences disappear upon normalization.
Thus the classical column load bound becomes
\begin{equation}
 O(I/J+H^{3/4}+D).                                 \label{eq:formula-21}
\end{equation}
For leaf size comparable to the original size up to a logarithmic
factor, $D$ is polylogarithmic in the leaf size. All wire, selected
load, precision, and degree estimates from Section
\ref{sec:forest-parity} therefore continue to hold. 

\subsection{Disjoint reduction for a fixed exceptional set}
We fix integers $2\leq t<D$ and set
\begin{equation}
 b=\min\left(n,\left\lfloor\frac{2(n-1)}{t+1}\right\rfloor\right).
                                                               \label{eq:formula-22}
\end{equation}
Every forest induced on $S$ has at most $b$ vertices of degree greater
than $t$, since its degree sum is at most $2(n-1)$. Denote this set
by $H_t(S)$. We enumerate all sets $B\subseteq V(G)$ of size at most
$b$. For fixed $B$, the subproblem counts
\begin{equation}
 B\subseteq S,\qquad G[S]\text{ acyclic},\qquad
 H_t(S)\subseteq B,                                \label{eq:formula-23}
\end{equation}
with any additional original selector pins. If $G[B]$ has a cycle
or a vertex of $B$ is pinned deleted, this count is zero.

Maintain an active induced graph $R=G[W]$ and a set $P\subseteq W$
of vertices pinned retained, initially including $B$ and the original
retained pins. Deleted pins are removed from $W$. If there is
$v\in W\setminus B$ of current degree $d_R(v)>D$, proceed as follows.
If $v\notin P$, create one branch deleting $v$. For every possible
\emph{exact} selected-neighbor set
\begin{equation}
 T\subseteq N_R(v),\qquad |T|\leq t,\qquad
 N_R(v)\cap P\subseteq T,                          \label{eq:formula-24}
\end{equation}
create a branch retaining $v$, pinning every vertex of $T$ retained,
and deleting every vertex of $N_R(v)\setminus T$.
If \eqref{eq:formula-24} has no solution, there is no retaining branch.
We use a deterministic rule to choose $v$.

\begin{lemma}[{Disjoint leaves and their density}]
Fix integers $2\le t<D$ and an exceptional retained set $B$ as in~\eqref{eq:formula-23}. The exact-neighbour branching procedure of this section partitions the solutions of \eqref{eq:formula-23}
among its leaves, with multiplicity one. At a leaf, each vertex
outside $B$ has input degree at most $D$ and the number of edges is
at most $ND$, where $N=|W|$. Its solutions are exactly the pinned
induced forests with degree cap $t$ on $W\setminus B$.
The number of leaves is at most
\begin{equation}
 2^{O\left((n/(D-t)+1)(t+2)\log(n+2)\right)}.       \label{eq:formula-25}
\end{equation}
The leaves can be streamed with polynomial working space.
\end{lemma}
\begin{proof}
A solution either omits $v$ or retains it. In the latter case,
$N_R(v)\cap S$ is its unique valid set $T$ in \eqref{eq:formula-24}, because
$v\notin B$ has degree at most $t$ in the final induced graph.
Pinning both $v$ and $T$ makes distinct retaining branches disjoint.
Future deletions cannot undo these pins. Conversely, every leaf
solution extends to an original solution by declaring removed
vertices deleted. The partition follows by induction.

A retaining branch deletes at least $d_R(v)-t>D-t$ active vertices;
none is already pinned retained by \eqref{eq:formula-24}. There can be at most
$n/(D-t)$ such steps on a path, while every path has at most $n$
steps in total. Each retaining step has at most $(n+1)^{t+1}$ choices.
Encoding the positions of these steps and their choices proves the
loose bound \eqref{eq:formula-25}, including paths ending in an empty subproblem.
Graph and pin descriptions along a depth-first path occupy polynomial
space; no family of leaves is stored.

At termination, degrees outside $B$ are at most $D$ in the whole
active graph, including neighbors in $B$. Edges with an endpoint
outside $B$ number at most $(N-|B|)D$. Edges entirely in $B$ number
at most $|B|-1$ when $B$ is nonempty, since $G[B]$ was checked acyclic.
For $B=\varnothing$ the simpler bound $ND/2$ applies. Both imply
$|E(R)|\leq ND$. The cap in \eqref{eq:formula-20} enforces every remaining
condition of \eqref{eq:formula-23}.
\end{proof}

\subsection{Removing multiplicities between exceptional sets}
Different $B$ can still represent the same induced forest. Let
$N_{j,h}$ be the number of original induced forests with $j$ retained
vertices and exactly $h$ vertices of degree greater than $t$.
For $0\leq k\leq b$, sum the leaf counts for all $k$-element $B$,
and let $c_{j,k}$ be their number of $j$-vertex solutions.
For a fixed solution $S$ with $|S|=j$ and $|H_t(S)|=h$, its admissible
guesses are exactly the sets $H_t(S)\subseteq B\subseteq S$ of size
$k$. Therefore
\begin{equation}
 c_{j,k}=\sum_{h=0}^{k}N_{j,h}\binom{j-h}{k-h}.     \label{eq:formula-26}
\end{equation}
Terms with impossible sizes are zero. For each $j$, recover in order
\begin{equation}
 \begin{split}
 N_{j,0}&=c_{j,0},\\
 N_{j,h}&=c_{j,h}-\sum_{r=0}^{h-1}N_{j,r}
                              \binom{j-r}{h-r}
       \quad(1\leq h\leq\min(b,j)).
 \end{split}                                                    \label{eq:formula-27}
\end{equation}
The total number of induced forests of size $j$ is
$\sum_{h=0}^{\min(b,j)}N_{j,h}$, since every such forest has $h\leq b$.
The triangular matrix in \eqref{eq:formula-26} has diagonal one, so no division or
precision loss occurs. Moreover $c_{j,k}\leq2^n\binom nk\leq2^{2n}$;
all reconstruction and accumulation use polynomially many bits.
Original selector pins preserve \eqref{eq:formula-26}, because for each surviving
$S$ all of its admissible $B$ are still enumerated.

\begin{proof}[Proof of Theorem~\ref{thm:general-fvs}]
The proof combines the sparse leaf cost with the exceptional-set and branch counts. Let $\lambda=\lceil\log_2(n+2)\rceil$, choose
$t=\max(2,\lceil\sqrt\lambda\rceil)$ and $D=t+\lambda^2$.
For sufficiently large $n$, \eqref{eq:formula-22} gives $b=O(n/\sqrt\lambda)$.
The number of exceptional guesses is bounded by
\begin{equation}
 \sum_{k=0}^{b}\binom nk
   =2^{O(n\log\lambda/\sqrt\lambda)}.              \label{eq:formula-200}
\end{equation}
For each guess, \eqref{eq:formula-25} gives $2^{O(n/\sqrt\lambda)}$ leaves;
the additive $(t+2)\log(n+2)=O(\lambda^{3/2})$ is absorbed for
sufficiently large $n$. Neither bound assumes the input sparse.

If a leaf has $N\leq n/\lambda$ vertices, enumerate its subsets and
check forest membership, pins, and degree caps directly. This costs
$2^{O(n/\lambda)}\poly(n+m)$. Otherwise
$\log N=\Theta(\log n)$, and the density bound $|E(R)|\leq ND$
gives $L_R=O(\log\lambda)$. The capped extension \eqref{eq:formula-21} applies:
$D=O((\log N)^2)$, so the residual event load at every pinned capped
vertex is below $H^{3/4}$ for sufficiently large $N$.
Edgeless leaves use the direct base case of Section~\ref{sec:forest-parity}.
For the other leaves, we use Theorem~\ref{thm:fvs} on the capped object family. Its cost is
\begin{equation}
 \poly(n+m)2^{O\left(
     N(\log\lambda)^2/\sqrt{\log N}\right)}
 \leq\poly(n+m)2^{O\left(
     n(\log\lambda)^2/\sqrt\lambda\right)}.         \label{eq:formula-201}
\end{equation}
All initial parity-matrix construction, greedy candidates, high-incidence pins,
count/order arrays, and integer recovery are already included in this
leaf bound. Multiplying by \eqref{eq:formula-200} and the number of leaves preserves
\eqref{eq:formula-28}. Handle bounded exceptional sizes directly.

Accumulate $c_{j,k}$ sequentially, using the original retained size
$j=N-\text{leaf deletion size}$; vertices removed before the leaf are
deleted and contribute no retained vertices. We apply \eqref{eq:formula-27} and reverse
the cardinality index to get feedback vertex counts. Streaming guesses,
branches, and leaf counters keeps polynomial space.

For recovery, find the first nonzero deletion-size count. We fix vertices
successively, count the branch deleting the next vertex, and choose
or skip that branch according to the desired rank. Arbitrary original
pins are supported by the same proof. There are $O(n)$ invocations of
the whole counter per recovered set, preserving \eqref{eq:formula-28}. Repeating for
successive ranks gives the stated enumeration bound.
\end{proof}

\section{Components, induced trees, and induced paths}
The forest identity can retain a component marker $T$ while using $V$ for
deletions. At a vertex use mass $1+(T-1)z_v$ and at an edge use
$(\bar\psi_u-\bar\psi_v)(\psi_u-\psi_v)-T\tau_u\tau_v$.
Deleted isolated vertices then have mass one. The separated polynomial is
\[
 \sum_{S:\,G[S]\text{ forest}}T^{c(G[S])}V^{n-|S|}.
\]
For an $N$-vertex leaf, evaluating $T=0,\ldots,N$ at $N$ additional bits
of precision recovers all component coefficients by finite differences.
The construction and precision proof appear in Appendix~\ref{sec:components}.
\begin{theorem}[{All-density size--component counts}]
\label{thm:components}
Under the Hypothesis~\ref{hyp:contraction}, for an arbitrary simple
graph all integers
\begin{equation}
 C_{d,c}=\#\{S\subseteq V(G):|V(G)\setminus S|=d,
                  \ G[S]\text{ a forest},\ c(G[S])=c\}
                                                               \label{eq:formula-31}
\end{equation}
can be computed deterministically in polynomial space and time
\begin{equation}
 \poly(n+m)2^{O\left(
       n(\log\log(n+4))^2/\sqrt{\log(n+2)}\right)}. \label{eq:formula-32}
\end{equation}
Arbitrary original selector pins are allowed. A set with prescribed
size and component count can be recovered or unranked within the
same bound; all such sets can be streamed with this delay.
\end{theorem}

\begin{proof}
We use the marked counter at each disjoint leaf. The binomial inversion is
applied independently at each component count. Appendix~\ref{sec:components}
gives the local mass events, degree and precision bounds, and recovery proof.
\end{proof}
\begin{corollary}[{Induced trees and induced paths}]
Assuming Hypothesis~\ref{hyp:contraction}, all size counts of induced trees and of
induced paths in an arbitrary simple graph, as well as a largest
such object, can be computed deterministically within \eqref{eq:formula-32} and
polynomial space. This includes containment of specified vertices,
exclusion of specified vertices, exact integer recovery and unranking.
Induced paths are counted by vertex set, with no orientation multiplicity.
\end{corollary}
\begin{proof}
For trees select $c=1$ in \eqref{eq:formula-31}; a single retained vertex is a tree,
while the empty set has $c=0$ and is excluded.
For paths impose degree cap two on every retained vertex. In this
case the exceptional set is known to be empty: use the disjoint
branching directly with $t=2$ and $D=2+\lambda^2$. Its leaf cost
and branch bound are within \eqref{eq:formula-32}, and every remaining input
vertex has degree at most $D$. The capped marked counter with
$c=1$ counts exactly connected acyclic graphs of maximum degree
two, namely paths including a single vertex. Selecting the largest
nonzero size gives deterministic optimization. Prefix-pin counts
recover the original vertex set; tracing from either endpoint
provides its path order. Prescribed vertices are ordinary pins.
\end{proof}

\section{Sparse companion counts}\label{sec:companions}
The same subset-index encoding supports directed Hamilton cycles,
edge-cardinality counts of unrooted spanning forests, and vertex-cardinality
counts of Steiner edge trees. We put $s=n(1+\log_2(1+m/n))$.

\begin{corollary}\label{cor:sparse}
Assuming Hypothesis~\ref{hyp:contraction}, these three counting problems
admit deterministic algorithms in
$\poly(n+m)2^{O(s/\sqrt{\log(s+2)})}$ time and polynomial space.
Hamilton and minimum Steiner witnesses can be recovered in the same bound.
Steiner counts support selector pins, forbidden edges, and required forest
edges. For fixed $C$ and $m/n\le(\log n)^C$, the time is
$\poly(n+m)2^{O(n\log\log n/\sqrt{\log n})}$.
\end{corollary}
\begin{proof}
Appendix~\ref{app:companions} gives the signed Hamilton identity, forest
coefficient recovery, and the selected-vertex Steiner polynomial, together
with all degree, precision, and recovery bounds.
\end{proof}

\section{Open problems}
The all-density reduction uses the sparsity of each retained induced forest.
Obtaining a vertex-subexponential bound for Hamilton counting on dense
graphs calls for a different density reduction. Exact counting of general
connected vertex sets is a second question, because cyclic selected graphs
introduce spanning-tree multiplicities. Within the present construction,
reducing the iterated-logarithm loss requires tighter filtering or encoding
budgets.

\label{end:maintext}
\section*{AI assistance}
OpenAI Codex was used to assist with drafting and revising the manuscript,
developing parts of the mathematical arguments, and checking derivations.

\bibliographystyle{splncs04}
\bibliography{refs}
\appendix
\section{Exact coefficient recovery}\label{app:recovery}
\subsection{Coefficient recovery without a large formal coefficient ring}
Let $f(X)$ have integer coefficients and degree at most $D$.
Values at $0,\ldots,D$ modulo $2^{p+D}$ determine all its power
coefficients modulo $2^p$. Indeed,
\begin{equation}
 f(X)=\sum_{j=0}^{D}\frac{\Delta^j f(0)}{j!}
                 X(X-1)\cdots(X-j+1),               \label{eq:formula-33}
\end{equation}
and $\Delta^j f(0)$ is divisible by $j!$ for an integer-coefficient
polynomial. Removing $2^{v_2(j!)}$, then inverting the odd part of
$j!$ modulo $2^p$, is valid because $v_2(j!)\leq D$.

For separate degrees $D_1,D_2$, apply this procedure successively
using precision $p+D_1+D_2$ at the evaluation grid.
Only polynomially many scalar contractions are used.

\section{Constructing the sparse-check shifts}\label{app:shifts}
\begin{lemma}[{Polynomial-time shift construction}]
\label{lem:bias}
Let $q=2^\ell\geq2$, $a\geq2$ be an integer,
$\mu\geq\max(1,\ell)$, and $d\geq8a^2\mu$.
A deterministic algorithm constructs $s_1,\ldots,s_d\in\mathbb F_2^\ell$
such that, for every nonzero $b$,
\[
 \left|\sum_{j=1}^d(-1)^{b\cdot s_j}\right|<d/a.
\]
Its bit complexity and space are polynomial in $q,d,a$.
For the parameters of source Lemma 6.1, this replaces the exhaustive
shift search by polynomial time in $m$ without changing its row,
support, or distance bounds.
\end{lemma}
\begin{proof}
This is an explicit conditional-expectation construction, using a
standard derandomization method~\cite{raghavan1988}; efficient small-bias
spaces are classical~\cite{naornaor1993}. The following argument specifies the constants and exact arithmetic at this interface.

We put $\epsilon=1/a$, $r=1+\epsilon$, and
$K=(r+r^{-1})/2$. After $j$ choices let
$S_b=\sum_{i\leq j}(-1)^{b\cdot s_i}$ and
\[
 \Phi_j=\sum_{b\ne0}(r^{S_b}+r^{-S_b}).
\]
For uniform next shift $s$, each nonzero character is balanced.
Therefore the average next potential is $K\Phi_j$. Choosing a shift
that minimizes it gives $\Phi_d\leq2(q-1)K^d$.
For $0<\epsilon\leq1/2$,
\[
 \log K\leq\epsilon^2/2,\qquad
 \log r\geq\epsilon-\epsilon^2/2,\qquad
 \epsilon\log r-\log K\geq\epsilon^2/4.
\]
Here logarithms are natural. Moreover
$\log(2(q-1))<(\mu+1)\log2\leq2\mu\log2<2\mu$
and $d\epsilon^2/4\geq2\mu$. Consequently
$2(q-1)K^d<r^{\epsilon d}$. If any $|S_b|\geq\epsilon d$,
its term alone is at least $r^{\epsilon d}$, a contradiction.

No real arithmetic or approximate comparison is needed. At step $j$
store the two integers
\[
 W_b^+=[a(a+1)]^j r^{S_b},\qquad
 W_b^-=[a(a+1)]^j r^{-S_b}.
\]
Their exponents of $a$ and $a+1$ are $j\mp S_b$, hence nonnegative.
Initially both are one. We write $A=(a+1)^2$, $C=a^2$.
A positive character multiplies $(W_b^+,W_b^-)$ by $(A,C)$,
and a negative character by $(C,A)$. Minimizing the next potential
is equivalent to minimizing
\[
 \sum_{b\ne0}(-1)^{b\cdot s}(W_b^+-W_b^-).
\]
All $q$ scores are the Walsh transform of the difference vector, with
zero at character zero. A fast exact transform uses $O(q\log q)$
integer additions per chosen shift. The stored integers have
$O(d\log(a+1))$ bits; there are $O(q)$ of them, and updates multiply
by integers of $O(\log(a+1))$ bits. All operations have polynomial bit
complexity and space, including the output list.

At the source interface set $a=200B\mu$,
$d=\min(m,320000B^2\mu^3)$, and take the least power of two $q$
with $qd\geq m$. If $q=1$, every shift is zero and there is no
nonzero character to constrain. Otherwise $d=320000B^2\mu^3
=8a^2\mu$, $q\leq m<2^\mu$, and $qd<2m$.
Thus all assumptions hold. In this nontrivial case $d,q\leq m$
and $a$ is polynomial in $m$, so the construction is polynomial in
$m$. Its bias is stronger than the source's required
$1/(200B\mu)$. The mixing and shortening argument then gives the stated support and distance bounds.
\end{proof}

\section{Polynomial-query selection by weighted conditional expectation}
\label{sec:conditional}
Weighted conditional expectations construct each separating row one position at a time.

We set $P_* = \min(P,m)$. Every nonzero violation surviving the initial
matrix has weight at least $m/P_*$. For $P\leq m$ this is the source
distance bound; for $P>m$ it follows from its integer weight being at
least one. Thus a random length-$P_*$ list still rejects every such
violation with probability at least $1/4$.

For current rows $A$, prefix row $h$, and $0\leq r\leq P_*$, define
the nonnegative integer moment
\begin{equation}
 W_r(A;h)=\sum_{\substack{(S,F)\in\Omega'\\Av=0,\ h\cdot v=0}}
                        (m-|v|)^r,\qquad
 W_r(A)=\sum_{\substack{(S,F)\in\Omega'\\Av=0}}(m-|v|)^r.
 \label{eq:moment}
\end{equation}
The pinned and degree-capped object family $\Omega'$ is permitted,
and $0^0=1$. The weight counts ordered $r$-tuples of coordinates at
which $v_e=0$, giving a direct nonnegative counting interpretation.
If $r$ random list positions remain, the expected number
accepted by the final appended row is exactly
\begin{equation}
 \mathcal E_r(A,h)=\frac{N(A)}2+
       \frac{2W_r(A;h)-W_r(A)}{2m^r}.             \label{eq:formula-34}
\end{equation}
Indeed a random position $(e,c)$ has character expectation
$1-|v|/m$, and the positions are independent. Expanding the indicator
$[h\cdot v=0]=(1+(-1)^{h\cdot v})/2$ proves \eqref{eq:formula-34}.

At the next position enumerate only its $2m$ choices. For each
$h'=h+c e_e$ compute $W_{r-1}(A;h')$ and choose a minimum.
All other terms in \eqref{eq:formula-34} are common to these children. This therefore
minimizes their conditional expectation, without computing any rational
number or requiring the unknown zero-violation count. Iterating for
$P_*$ positions yields a row at least as good as the initial random-list
expectation. Repeating at most $3H$ times gives exact separation by the
same $(3/4)^{3H}2^H<1$ argument as before.
There are at most $6HP_*m$ weighted oracle calls, plus an initial
unweighted count if early stopping is used. The last-position moment
has $r=0$ and is already the unweighted count of the chosen row.

\begin{lemma}[{Weighted ordinary-network oracle}]
\label{lem:moment}
For $m\leq n^{33/32}$ and sufficiently large $n$, put $L=1+\log_2(1+m/n)$. Every weighted moment $W_r(A;h)$ or $W_r(A)$ in~\eqref{eq:moment}, for $0\le r\le P_*$ and for the pinned, degree-capped object family specified there, has an ordinary scalar-network representation with the same
asymptotic root scale
$s=nL[L+\log\log(n+4)]$ as \eqref{eq:formula-46}. Its complete evaluation cost
under D.1, including public arrays and heavy-variable pins, is
$\poly(n+m)2^{O(s/\sqrt{\log(n+2)})}$ with polynomial working space.
Consequently the separating rows require polynomially many weighted calls within \eqref{eq:formula-45}.
\end{lemma}
\begin{proof}
The weighted oracle encodes the moment as labeled selector slots while preserving whole-edge cancellation. In every object, $F\subseteq E(G[S])$. The integer violation weight is
therefore $|v|=\sum_{uv\in E}z_uz_v-\sum_e f_e$. We add the factor
\begin{equation}
 Q(z,f)^r=\left(m-\sum_{uv\in E}z_uz_v+\sum_e f_e\right)^r.
                                                               \label{eq:formula-35}
\end{equation}
It is a function only of selectors and whole-edge events, so it
preserves the coefficientwise forest cancellation. On genuine objects
it equals $(m-|v|)^r\geq0$. The prefix row $h$ is represented by one
ordinary parity lift, in exactly the same way as any candidate row.
Throughout the selection process the total parity row count is at
most $6H$ and its support bounds remain $D_0$ and $P_*$.

We represent \eqref{eq:formula-35} by $r$ labeled slots. Each slot chooses the constant
$m$, one term $-z_uz_v$, or one term $+f_e$. Constant choices have no
ports; other choices have at most two classical predicate ports and
no physical atoms or degree events. We put all slots in one additional
factor row. Its entry checks that a slot selects either no term or
exactly one complete term, and multiplies the resulting coefficients.
Classical columns enforce all positive predicates. After pins, collect
constant terms within each slot and delete or shorten the remaining
predicates; this does not increase incidence. Distinct slots retain
their labels, so all ordered products in $Q^r$ occur with their correct
multiplicities. This moment polynomial is over the integers: duplicate
predicates retain their integer multiplicities. The moment row uses the subset-index encoding of Section~\ref{sec:sparse-global}.

We use precision
\[
 P_0=H+P_*\mu+2.
\]
Every required nonnegative count is less than $2^H m^r<2^{P_0}$.
Thus the least nonnegative residue is the exact integer used in the
comparison. The parity-lift order cutoff must be recomputed as
$P_0+a-1$, where $a$ is the current row count. For the stated density
range, $B=O(n^{1/32})$, hence
$P_*\mu=O(n^{1/16}\log^3(n+2))=o(n)$ and $P_0=O(H)$.
There are only $2r=o(H)$ additional selected ports. The original
$O(H)$ selected-port and precision budgets are preserved.

Raw incidence is now
\[
 I'=O(n+m+H\tau+P_*m)
    =O(n^{35/32}\log^4(n+2)).
\]
The subset entropy bound still gives
$O(H[1+\log(1+I'/H)])=O(s)$ wires, since
$\log(1+I'/H)=O(L+\log\log(n+4))$.
With the same $J=\lceil H^{1/4}\rceil$ groups, the additional row
has wire degree $O(J+P_*\log(I'+2))$ by its selected-port bound.
We pin variables of incidence greater than $H^{3/4}$ and rebalance the
classical columns as before. The number of pinned variables and the
maximum remaining degree are bounded respectively by
\begin{equation}
 O(n^{11/32}\log^4(n+2)),\qquad
 O(n^{27/32}\log^4(n+2)).                         \label{eq:formula-36}
\end{equation}
For capped FVS leaves, add the same $O(D)$ residual event load from
\eqref{eq:formula-21}; $D$ is polylogarithmic and is absorbed. Since $27/32<9/10$,
the degree condition of Theorem D.1 holds for sufficiently large $n$.
The extra factor row adds $O(J)$ buckets, so the array exponent is $O(\sqrt H\log(H+2))$. Both this exponent and the
pin-branch exponent in \eqref{eq:formula-36} are absorbed by $O(s/\sqrt{\log n})$.
All local entries, including constant powers and invalid-index checks,
are computable in polynomial bit complexity and space.

The number of moments is polynomial in $n+m$, each is evaluated
sequentially, and their descriptions have polynomial length. The oracle and selection costs follow. Small sizes and the denser regime retain
the direct-enumeration treatment of Theorem~\ref{thm:fvs}.
\end{proof}

\section{Component markers and recovery}

\label{sec:components}

The counter in Section~\ref{sec:dense-fvs} can retain the exact
number of components without increasing its main exponent. The
component variable is already present in the forest identity of
Caracciolo et al.~\cite{forest}, equation (13). The component-counting interface must
ensure that deleted isolated vertices contribute no component marker,
and must recover two statistics using the source's permitted
single-variable coefficient ring.

\subsection{Marked mass choices with deleted vertices removed}
We write $T$ for the component marker and $V$ for the deletion marker.
For a fixed selector vector $z$ with retained set $S$, put
\begin{equation}
 a_v(T,z)=1+(T-1)z_v,\qquad
 Q_e(T)=(\bar\psi_u-\bar\psi_v)(\psi_u-\psi_v)-T\tau_u\tau_v.
                                                               \label{eq:formula-37}
\end{equation}
The multivariate identity is
\begin{equation}
 \Top\prod_v(1+a_v(T,z)\tau_v)
       \prod_{e=uv}(1+w_e z_uz_v Q_e(T))
   =\sum_{F\subseteq E(S),\ F\text{ acyclic}}
        T^{|S|-|F|}\prod_{e\in F}w_e.              \label{eq:formula-38}
\end{equation}
Indeed, after fixing $z$, deleted vertices have no incident edge
terms and their mass integral is one. On $S$, source equation (13)
has uniform parameters $t=T,u=-T$, hence one factor $T$ per forest
component. Thus \eqref{eq:formula-38} follows over the integers coefficientwise
in the independent edge variables. It also holds for $S=\varnothing$,
whose component exponent is zero.

We keep the whole-edge macro convention of Section~\ref{sec:forest-parity}.
We multiply each quartic edge monomial's coefficient by $T$. For a mass
term at a pinned retained vertex, multiply its coefficient by $T$;
at a pinned deleted vertex, leave the coefficient one. A mass term
at an unpinned vertex sends one additional \emph{mass-choice event}
to that vertex's classical column. The event records selection of a mass monomial. It is defined for every selector value, and degree caps count the separate edge events.

Let $g_v$ record whether that mass term was selected. An unpinned
vertex required retained by an edge macro or parity predicate
contributes $T^{g_v}$ at its column. If it has no such requirement,
its column sums the two choices with factor
\begin{equation}
 V+T^{g_v}.                                       \label{eq:formula-39}
\end{equation}
Pinned deleted vertices contribute $V$ as before. Every nonzero
full assignment has exact physical-atom coverage. Consequently a
vertex without its mass term is incident to a selected edge term
and is required retained; the apparently unrestricted alternative
in \eqref{eq:formula-39} cannot create a spurious deleted vertex in that case.
This describes polynomial-time entries using only the ports already
incident to their row or column.

There are at most $N$ additional selected mass-event ports on an
$N$-vertex leaf. The total selected-port bound becomes
$6N+2K_0=O(H)$, and input incidence increases by at most $N$.
Each mass term still consumes its owner atom, so the graph-row bound
has the same bound. Each free selector column receives one mass occurrence; pinned selectors have fixed mass values. The degree caps
count edge events. The sign calculation and degree bounds are those of
Sections~\ref{sec:forest-parity} and \ref{sec:dense-fvs}.

For a parity matrix $A$, the resulting polynomial is exactly
\begin{equation}
 F_A(T,V)=\sum_{(S,F)\in\Omega':\ Av(S,F)=0}
                  T^{|S|-|F|}V^{N-|S|},           \label{eq:component-1}
\end{equation}
where $\Omega'$ includes the current pins and degree caps.
The greedy separation of Lemma~\ref{lem:separation} is applied at $T=1,V=1$. Its completed
matrix rejects every incorrect object, independently of $T,V$.
Hence its final polynomial is
\begin{equation}
 F_A(T,V)=\sum_{S:\ G[S]\text{ forest, pins/caps satisfied}}
                     T^{c(G[S])}V^{N-|S|}.         \label{eq:component-2}
\end{equation}
The same separating matrix serves every component value.

\subsection{Two statistics with one formal variable}
The degree in $T$ is at most $N$ even before cancellation in the
signed-cover expansion. Every selected mass contributes at most
one $T$ and consumes two physical atoms; every quartic edge
contributes one $T$ and consumes four. There are only $2N$ atoms.
The deletion degree is also at most $N$.

Let $p=H+1$, with $H$ defined for the leaf as in \eqref{eq:formula-4}. For each
integer $i=0,\ldots,N$, contract the marked network at $T=i$ over
\begin{equation}
 (\mathbb Z/2^{p+N}\mathbb Z)[V]/(V^{N+1}).        \label{eq:component-3}
\end{equation}
The lifted parity truncation uses this increased precision, so its
total retained order is $p+N+a-1=O(H)$. Both the precision and the one formal truncation length
remain $O(s)$ at the source root scale. Local entries compute powers of $i$ modulo $2^{p+N}$, using $p+N$ bits of precision.

For each deletion coefficient $d$, apply the integer-polynomial
Newton recovery \eqref{eq:formula-33} to
$[V^d]F_A(T,V)$. Its degree is at most $N$, and the $j$th difference
is divisible by $j!$. Removing the power of two in $j!$ loses at
most $N$ bits, and the odd factor is invertible. Thus the values
modulo $2^{p+N}$ determine every power coefficient modulo $2^p$.
All final coefficients count vertex sets and are at most $2^N<2^p$,
so their least nonnegative residues are the exact integers.

Only $N+1$ contractions are used, each in the source's permitted
single-variable ring. 
The value table and coefficient table require polynomial space.

\paragraph{Proof details for Theorem~\ref{thm:components}.}\begin{proof}
We apply the same exceptional-set guesses and disjoint sparse branches
as Theorem~\ref{thm:general-fvs}. At each leaf compute \eqref{eq:component-2}
using \eqref{eq:component-3}, or directly enumerate if the leaf is small. This
adds only a polynomial number of source contractions at the same
scale. Vertices removed before reaching a leaf are deleted, so the
retained size is $N-d_{\rm leaf}$ and its component count is exactly
the component count in the original graph.

For each fixed retained size $j$ and component count $c$, the
exceptional-set sum is
\begin{equation}
 c_{j,k,c}=\sum_{h\leq k}N_{j,h,c}\binom{j-h}{k-h}. \label{eq:component-4}
\end{equation}
We apply \eqref{eq:formula-27} separately for every $c$ and sum over $h$. The component-count array has polynomially many entries, and its nonnegative counts have polynomial bit length. The outer exponents are those of Theorem~\ref{thm:general-fvs}. For a prescribed pair of statistics, prefix-pin counts
give unranking exactly as before, using $O(n)$ full-count calls.
\end{proof}

\section{Sparse companion constructions}\label{app:companions}
\subsection{Hamilton cycles and the full forest polynomial}
For an arc $u\to v$, make two optional terms on the physical atoms
$\mathrm{out}_u,\mathrm{in}_v$. The grade-zero term has weight one.
The grade-one term has weight minus the sign needed to put
$\mathrm{out}_u\mathrm{in}_v$ into canonical atom order.
At every vertex require equal incoming and outgoing grades, and at a
fixed root require grade zero. Every surviving cover is a directed
cycle cover with constant grade on each cycle. If $A$ is the active
vertex set, the exterior sign is $\operatorname{sgn}(\pi_A)$ and
the explicit signs supply $(-1)^{|A|}$. Thus each active cycle has
weight $-1$, and a cover with $c$ cycles contributes $(1-1)^{c-1}$.
The complete signed sum is exactly the integer Hamilton count.

For forests, use the $2n$ Grassmann atoms
$\bar\psi_1,\psi_1,\ldots,\bar\psi_n,\psi_n$, set
$\tau_v=\bar\psi_v\psi_v$, and let $\Top$ extract the coefficient of
$\tau_1\cdots\tau_n$ in canonical order. The imported forest identity is
\begin{equation}
 \Top\prod_v(1+\tau_v)\prod_{uv\in E}(1+UQ_{uv})
       =\sum_{F\text{ forest}}U^{|F|},                \label{eq:companion-forest}
\end{equation}
where $Q_{uv}=(\bar\psi_u-\bar\psi_v)(\psi_u-\psi_v)-\tau_u\tau_v$.
We replace each edge factor by the product of its five elementary
monomial factors, and each vertex factor by its single mass factor.
This is an integer identity: the even monomials commute, each has
square zero, and all products of two distinct summands within
$Q_{uv}$ cancel, since its quadratic part is decomposable and its
quartic part overlaps it. Equivalently, the product is the finite
exponential of $UQ_{uv}$, which equals $1+UQ_{uv}$.
Each elementary term has two or four physical atoms and grade one.
There are $O(n+m)$ ports, so \eqref{eq:formula-3} applies. The degree in $U$ of
each selected-term contribution is at most $n$; cancellation gives
the usual forest degree at most $n-1$.

Coefficient recovery uses Appendix~\ref{app:recovery}.
Both the Hamilton count and every coefficient of the forest polynomial
are bounded by
\begin{equation}
 \sum_{j=0}^{n}\binom mj\leq\binom{m+n}n
                  \leq[e(1+m/n)]^n.                 \label{eq:companion-1}
\end{equation}
For Hamilton cycles the simpler product-of-outdegrees bound also works.
An additional factor $2^n$ covers all vertex selections below.
Thus $p=O(s)$ suffices, and the $O(n)$ extra interpolation bits
preserve $P=O(s)$.

\paragraph{Hamilton and forest counts.}
\label{thm:sparse-global}
Under Hypothesis~\ref{hyp:contraction}, directed Hamilton cycles and all edge-cardinality
counts of unrooted spanning forests in an $n$-vertex, $m$-edge graph can
be computed deterministically in
\begin{equation}
 \poly(n+m)\,2^{O(s/\sqrt{\log(s+2)})}              \label{eq:companion-2}
\end{equation}
time and polynomial space, with $s$ as in \eqref{eq:formula-1}.
The bound includes integer recovery and Hamilton witness recovery.
Undirected Hamilton cycles on at least three vertices follow by
bidirecting the edges and dividing the integer answer by two.

\begin{proof}
The scalar network satisfies the contraction premise at the claimed scale. For sufficiently large $n$, the maximum degree in the preceding lemma
is at most $\lceil s^{9/10}\rceil$. The number of vertices and the
description length are polynomial, and local entries take polynomial
time and space. We use Hypothesis~\ref{hyp:contraction} with $\ell=1$ and the precision
just proved. The array exponent $O(\sqrt n\log n)$ and the polynomial
number of interpolation calls are absorbed by \eqref{eq:companion-2}. We enumerate
arrays and calls sequentially. Bounded exceptional sizes use direct
enumeration. Hamilton recovery deletes an arc whenever the remaining
count is nonzero, yielding a single cycle in at most $m$ further calls.
All subgraphs retain the original budget as an upper bound.
\end{proof}

In particular, \eqref{eq:companion-2} is vertex-subexponential when
$\log(1+m/n)=o(\sqrt{\log n})$. For $m/n\leq(\log n)^C$,
fixed $C$, it is $2^{O(n\log\log n/\sqrt{\log n})}\poly(n+m)$.
For $m=O(n)$ it retains $2^{O(n/\sqrt{\log n})}$.
The dense worst case $m=\Theta(n^2)$ still has $s=\Theta(n\log n)$;
the scale is $\Theta(n\log n)$.

\subsection{Steiner counting, high-degree selectors, and original witnesses}
Let a nonempty terminal set $T$ be fixed. Introduce selector bits $z_v$,
pin $z_v=1$ for $v\in T$, give each selected vertex weight $V$ and each
unselected vertex weight $U$, and replace an edge weight in \eqref{eq:companion-forest}
by $Uz_uz_v$. The exact polynomial is
\begin{equation}
 B(U,V)=\sum_{S\supseteq T}U^{n-|S|}V^{|S|}
                 \sum_{F\subseteq E(S),\ F\text{ forest}}U^{|F|}.
                                                               \label{eq:companion-3}
\end{equation}
Its coefficient $[U^{n-1}V^j]$ counts precisely the $j$-vertex
Steiner trees as pairs $(S,F)$, each once. 

Each selected elementary edge term now has two positive predicate
ports, requiring $z_u=z_v=1$, in addition to its physical atoms.
A classical column tensor contributes $V$ for a vertex touched by
such a port and $U+V$ for an untouched unpinned vertex. Fixed selector
weights are scalars. At most $n$ elementary terms can be selected,
so the total number of selected physical and classical ports is at
most $4n$. The entropy bound \eqref{eq:formula-3} and the row-load bound survive.
Classical columns require a separate degree argument.

First consider $m\leq n^{17/16}$. We pin every unpinned selector whose
original graph degree exceeds $n^{3/4}$, in all possible ways. There
are at most $2m/n^{3/4}=O(n^{5/16})$ such vertices, and these branches
partition the selected sets. We remove terms that require a pinned zero
and remove predicates already fixed to one. Each remaining classical
vertex has $O(n^{3/4})$ input incidences. Greedily balancing their
incidence weights among $J$ columns gives raw load
\begin{equation}
 O((n+m)/J+n^{3/4})=O(n^{13/16})                  \label{eq:formula-47}
\end{equation}
per column. Subset-index wire degree never exceeds raw incidence count.
Thus all classical columns satisfy the source degree condition as well.

The polynomial in \eqref{eq:companion-3} has separate degrees at most $2n$ and $n$
even before cancellation. We use \eqref{eq:formula-33} with $p+3n$ bits at the scalar
grid; \eqref{eq:companion-1} with the extra $2^n$ factor bounds every final coefficient.
The high-degree branches cost $2^{O(n^{5/16})}$, absorbed by \eqref{eq:companion-2}.
For $m>n^{17/16}$, enumerate vertex subsets and compute the spanning-tree
count of each induced subgraph by a Laplacian cofactor. This costs
$2^n\poly(n+m)$ time and polynomial space, already within \eqref{eq:companion-2}
because $s=\Omega(n\log n)$ in this density range. This fallback
ensures the following statement applies to every simple graph.

\paragraph{Steiner edge-tree counts.}
All vertex-cardinality Steiner tree counts, the minimum cardinality,
and a minimum tree witness satisfy \eqref{eq:companion-2}, deterministically in
polynomial space under Hypothesis~\ref{hyp:contraction}.
The same bound holds with additional selector pins, forbidden edges,
and at most $n-1$ required forest edges.

\begin{proof}
The preceding construction proves counting. A required edge contributes the
factor $UQ_{uv}$, whose five alternatives form one
required factor, handled by the additional balanced rows. More than
$n-1$ required edges, or a required cycle, gives zero. For dense fallback,
required acyclic edges can be contracted before applying the cofactor
formula to each selected graph, with parallel-edge multiplicities retained.

We choose the smallest nonzero cardinality coefficient. We pin original
vertices one at a time while retaining a nonzero coefficient. Once the
selected set is fixed, delete edges while retaining a nonzero count.
The final edge-minimal connected graph is a tree. This uses at most
$O(n+m)$ complete count queries and returns original vertices and edges.
Unranking and sampling can similarly use original selector and edge
inclusion/exclusion queries.
\end{proof}

\paragraph{Finite checks.} The algebraic and reduction layers were checked
on every nonisomorphic simple graph with at most ten vertices.

\end{document}

%% file: technical_overview_body.tex
\begin{figure}[H]\centering
\begin{tikzpicture}[node distance=5mm, every node/.style={font=\small}, box/.style={draw,rounded corners,align=center,text width=.83\linewidth,inner sep=5pt}]
\node[box] (a) {Forest identity: signed expansion with exact cancellation};
\node[box,below=of a] (b) {Selected-port compression: control Boolean wires and signs};
\node[box,below=of b] (c) {Whole-edge parity filtering: a density-sensitive leaf counter};
\node[box,below=of c] (d) {High-degree branching: disjoint sparse leaves for each guess};
\node[box,below=of d] (e) {Binomial inversion: recover every FVS cardinality count};
\draw[->] (a)--(b);\draw[->] (b)--(c);\draw[->] (c)--(d);\draw[->] (d)--(e);
\end{tikzpicture}
\caption{The counting reduction. Compression and filtering construct the leaf oracle; branching and inversion extend it to arbitrary density.}\label{fig:overview}
\end{figure}
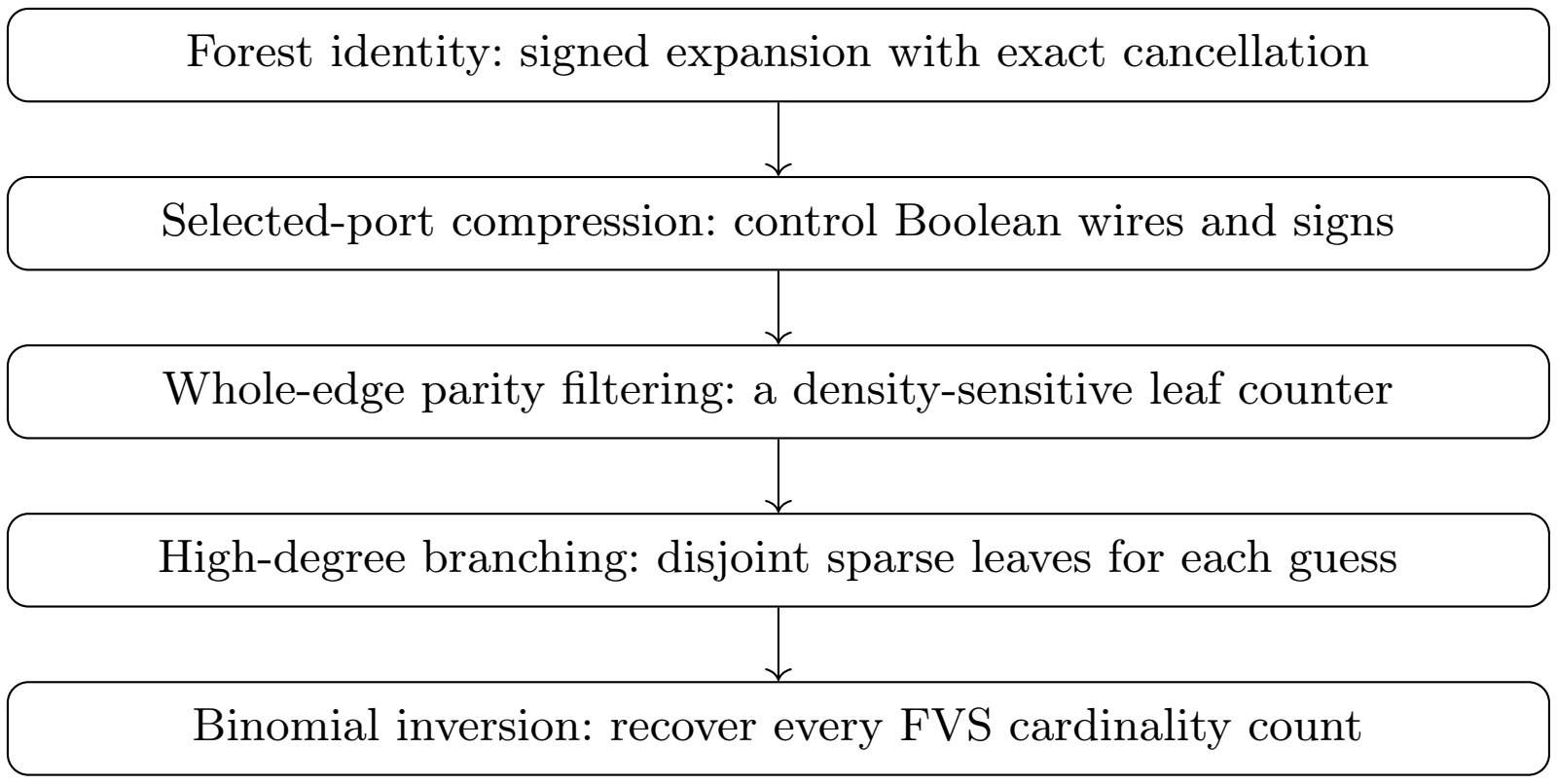

The algorithm separates the algebra of acyclicity from the cost of representing a graph. A forest identity supplies exact cancellation, while a sequence of encodings makes that identity small enough for the contraction theorem~\cite[Theorem D.1]{base}. The final reduction handles graph density outside the algebraic calculation. Figure~\ref{fig:overview} summarizes these roles; the details below explain why each arrow is needed.

\paragraph{A forest identity supplies the counting object.}
The starting point is the Grassmann identity in Section~\ref{sec:forest-identity}. Its top coefficient is a generating polynomial for forests, with one variable recording the number of selected edges. The signs cancel cyclic edge selections. Feedback vertex counting asks for induced forests: after choosing retained vertices, every edge between them must be present. Thus the desired object combines a vertex selection with an edge selection, and the two must agree. Keeping this distinction through the encoding is essential for recovering integer cardinality counts.

\paragraph{Compression pays for selected incidences.}
A direct incidence network has a wire for each graph incidence, so its size can grow with the number of edges. The algebraic expansion, however, uses only a bounded number of atoms at each selected term and at most linearly many selected terms. Subset indices encode these selected ports inside groups. Their entropy yields the density-sensitive scale $n(1+\log(1+m/n))$. The Grassmann sign is then factored into ordinary scalar tensors. This step lets the external contraction algorithm operate on Boolean wires with scalar entries; the sign bookkeeping stays inside the local tensors. The wire count, maximum degree, and cost of evaluating an entry are tracked together.

\paragraph{Parity checks enforce whole-edge agreement.}
The compressed expansion contains auxiliary choices before cancellation. A parity test must refer to a whole graph edge, including both endpoint requirements, so that all monomial realizations of that edge receive the same test. Coefficientwise cancellation then preserves the forest identity within each tested object class. Sparse initial checks exclude low-weight discrepancies, and additional checks separate the remaining positive object counts. A weighted conditional-expectation oracle chooses these checks with polynomially many contraction queries. Its role is algorithmic as well as algebraic: it includes the cost of selecting the filter, the precision of its counts, and the local-entry cost of its weighted queries. Section~\ref{sec:forest-parity} assembles these ingredients into the density-sensitive leaf counter.

\paragraph{High-degree branching makes the leaf counter useful on dense graphs.}
Every induced forest has few vertices whose retained degree exceeds a threshold $t$. The outer algorithm guesses a set $B$ containing them. Outside $B$, a retained vertex has at most $t$ retained neighbours. At a vertex with large current degree, a retaining branch specifies its exact retained-neighbour set and deletes all other neighbours; a deleting branch removes the vertex. Exact neighbour choices partition the solutions for this fixed $B$. The process ends with a graph of bounded density, where the compressed counter applies with degree caps. Choosing $t$ of order $\sqrt{\log n}$ and the branching threshold of order $(\log n)^2$ keeps both the exceptional-set guesses and the leaf count within the final exponent.

\paragraph{Binomial inversion removes overlap between guesses.}
The branches for one $B$ are disjoint, but a forest can admit several exceptional-set guesses. A size-$j$ forest with $h$ genuinely high-degree retained vertices occurs in exactly $\binom{j-h}{k-h}$ guesses of size $k$. Summing the leaf counts by $(j,k)$ therefore gives a triangular binomial transform of the desired $(j,h)$ counts. Its diagonal entries are one. Successive integer subtraction recovers the distribution without division, and reversing the retained-size index gives the FVS distribution. This is the final recovery step in the proof of Theorem~\ref{thm:general-fvs}.

\paragraph{The same interface supports the stated applications.}
Pinned vertices are carried through the branches and the leaf network. Counting the two possibilities for the next vertex gives minimum-solution recovery and unranking. Component markers retain the statistics needed for induced trees and paths. Every outer guess, branch, and contraction call is processed sequentially; polynomial working space follows from storing one recursion path and polynomially many exact accumulators. The appendices supply the coefficient recovery, shift construction, and weighted-query details used by this proof skeleton.